\documentclass[11pt]{article}
\usepackage[utf8]{inputenc}

\usepackage[T1]{fontenc}
\usepackage{lmodern}

\usepackage{fullpage}

\usepackage{indentfirst}
\usepackage[indent]{parskip}

\usepackage{amsmath}
\usepackage{amssymb}
\usepackage{amsthm}
\usepackage{mathtools}
\usepackage{thmtools}

\usepackage[
    pdftex,
    pdfauthor={},
    pdftitle={Efficient Multi-Agent Delegated Search}
]{hyperref}
\hypersetup{colorlinks=true, linkcolor=blue, citecolor=blue, urlcolor=blue}
\usepackage[nameinlink,noabbrev,capitalize]{cleveref}

\usepackage{enumitem}
\setlist{listparindent=\parindent, parsep=\parskip}

\usepackage{bbm}
\usepackage[numbers]{natbib}

\usepackage[framemethod=tikz]{mdframed}
\mdfsetup{backgroundcolor=none,skipabove=4pt,skipbelow=4pt,linewidth=1pt,bottomline=false,topline=false,rightline=false}

\theoremstyle{plain}
\newmdtheoremenv{theorem}{Theorem}[section]
\newmdtheoremenv{proposition}[theorem]{Proposition}
\newmdtheoremenv{lemma}[theorem]{Lemma}
\newmdtheoremenv{corollary}[theorem]{Corollary}
\newmdtheoremenv{conjecture}[theorem]{Conjecture}

\theoremstyle{definition}
\newmdtheoremenv{definition}[theorem]{Definition}

\theoremstyle{remark}
\newmdtheoremenv{remark}[theorem]{Remark}
\newmdtheoremenv{example}[theorem]{Example}

\newcommand{\R}{\mathbb{R}}
\newcommand{\E}{\operatorname*{\mathbb{E}}}

\renewcommand{\O}{\operatorname{\mathcal{O}}}

\newcommand{\union}{\,\cup\,}

\newcommand{\mathcaps}[1]{{\normalfont \textsc{#1}}}
\newcommand{\princ}{\operatorname{\mathcaps{Principal}}}
\newcommand{\opt}{\operatorname{\mathcaps{Opt}}}

\begin{document}

\title{Efficient Multi-Agent Delegated Search}
\author{
    Curtis Bechtel \\
    Department of Computer Science \\
    University of Southern California \\
    \texttt{bechtel@usc.edu} \\
    \and
    Shaddin Dughmi \\
    Department of Computer Science \\
    University of Southern California \\
    \texttt{shaddin@usc.edu} \\
}
\date{\vspace{-4ex}}

\maketitle

% TLDR: We consider a few related models of multi-agent delegation, and study mechanisms that approximate the principal's first-best expected utility.

\begin{abstract}
    Consider a \emph{principal} who wants to search through a space of stochastic solutions for one maximizing their utility. If the principal cannot conduct this search on their own, they may instead delegate this problem to an \emph{agent} with distinct and potentially misaligned utilities. This is called delegated search, and the principal in such problems faces a mechanism design problem in which they must incentivize the agent to find and propose a solution maximizing the principal's expected utility. Following prior work in this area, we consider mechanisms without payments and aim to achieve a multiplicative approximation of the principal's utility when they solve the problem without delegation.
    
    In this work, we investigate a natural and recently studied generalization of this model to multiple agents and find nearly tight bounds on the principal's approximation as the number of agents increases. As one might expect, this approximation approaches $1$ with increasing numbers of agents, but, somewhat surprisingly, we show that this is largely not due to direct competition among agents.
\end{abstract}

%%% Local Variables:
%%% mode: latex
%%% TeX-master: "main"
%%% End:

\section{Introduction}

``If you want something done right, you have to do it yourself'' may be little more than a catchy cliche, but it hints at an important idea in economic decision-making: when a \emph{principal} delegates a task to untrusted \emph{agents}, misaligned interests can lead to suboptimal outcomes. If this principal lacks the resources or ability to complete their own task, then they are faced with a mechanism design problem in which they want to select a delegation mechanism that optimizes their expected utility. In this paper, we aim to help the principal by finding multi-agent delegation mechanisms with competitive multiplicative approximations of what the principal could achieve on their own.

Consider the following scenario that helps to illustrate and motivate our particular model and results. Take the perspective of a committee within a governmental body that funds scientific research through grants. You are tasked with allocating a fixed amount of resources for a single research project that benefits the nation's long-term interests, and there are several research groups from which you can receive, evaluate, and approve proposals. You are confident in your ability to evaluate proposals, but recognize that each research group has its own interests that may be misaligned with the nation's. You must try to design a grant proposal mechanism that motivates research groups to propose research projects that are most beneficial to the nation.

More broadly, we consider models in which the principal faces a stochastic optimization problem where they have to find a solution maximizing the expected value of some objective function. The task of searching for solutions to this problem is then delegated to a fixed group of agents, who each have distinct utility functions. Agents propose solutions to the principal, and the principal picks a single winner who receives utility for their proposal. We focus on models of delegation in which the principal's mechanism can not make outcome-contingent payments, representing situations in which players are confined to a fixed-price contract or are not legally allowed to make transfers of value for specific outcomes.

Finally, in contrast to designing optimal mechanisms, we build on recent delegation research \cite{kleinberg2018delegated,bechtel2021probing,bechtel2022pandora,hajiaghayi2023agents} in which the principal aims for a multiplicative approximation of their first-best expected utility, i.e. their expected utility when the problem is not delegated (alternatively, their utility when they delegate to agents with identical interests). This approximation factor, which can be called the \emph{delegation gap}, tells the principal what fraction of their optimal utility they are guaranteed while delegating to arbitrary untrusted agents.

\subsection{Overview of Our Models}

The delegation model of primary interest in this paper is \emph{strategic multi-agent delegation} (\cref{def:strategic}) as originally defined by Hajiaghayi et al.\ \cite{hajiaghayi2023agents}. This consists of a principal who wants to find a solution maximizing their utility over a stochastic solution space, and they must delegate the search process to $k$ agents. Each agent is given a finite number of \emph{elements} from which they can sample an \emph{outcome} representing a solution along with the principal's and agent's utilities. The number of agents, number of elements per agent, and the distributions of outcomes are specified as part of the instance and common knowledge to all players.

In this game, the principal starts by committing to some mechanism through which agents can communicate with the principal. Each agent then samples outcomes from their elements and sends a signal to the principal. The principal transforms these signals into an outcome which is conditionally accepted as the winner. If this outcome was not sampled by any agent, then the principal detects this ``lie'' and rejects it, so all players get no utility. Otherwise, the principal and winning agent each get the utility specified by that outcome, and all other agents get nothing.

One challenge posed by this model is the complexity of analyzing agents' equilibrium strategies, how these equilibria are affected by the choice of mechanism, and how they affect the principal's utility. We also study a simplified model in which agents are assumed to act adversarially against the principal. More specifically, \emph{adversarial multi-agent delegation} (\cref{def:adversarial}) is the same as the strategic model above, except that we do not define agents' utilities and all agents instead aim to minimize the principal's expected utility subject to maintaining a positive probability of winning.

We will see that in the absence of symmetry conditions on the agents, it is difficult to make any nontrivial approximation guarantees in either model. One attempt to get around this involves the (strategic or adversarial) \emph{shuffled multi-agent delegation} model (\cref{def:shuffled}), in which elements are randomly distributed among agents. Specifically, there is a known pool of elements, each of which is given to a uniformly random agent, and the game proceeds from there as usual. Note that the principal does not learn which agent received which elements, and their expected utility is measured with respect to the random allocation of elements to agents. A perhaps more straightforward way of enforcing symmetry is the (strategic or adversarial) \emph{agent-symmetric multi-agent delegation} model, in which all agents are given access to equivalent sets of elements.

For these models, we consider three natural classes of mechanisms. First are \emph{single-proposal mechanisms} (\cref{def:single-proposal}), originally defined for this multi-agent context in \cite{hajiaghayi2023agents}, in which the principal announces restricted sets of outcomes that they would accept, each agent proposes a single outcome to the principal, and the principal picks a winner from among them. This is perhaps one of the most natural classes of mechanisms, and can be seen implemented, for example, in the form of research grants. Specifically, an agency sets out criteria of proposals that they would be willing to accept and then receives a single proposal from each research group. Importantly, it is known that single-proposal mechanisms perform at least as well as any other kind of mechanism (\cref{thm:single-proposal}).

As a special case of single-proposal mechanisms, we consider \emph{threshold mechanisms} (\cref{def:threshold}), in which the principal's acceptable sets of outcomes are defined by thresholds on their utility. Focusing on this restricted class is beneficial for vastly simplifying the complexity of describing an individual mechanism, reducing the space of mechanisms to something more manageable, and being more intuitive to a potential implementor.

Although we spend little time on it, we also define a class of direct-revelation mechanisms called \emph{Myerson-type mechanisms} (\cref{def:myerson-type}) that may be of interest for future work to expand on. As the name implies, this class is inspired by Myerson mechanisms from auction theory. In it, the principal declares a virtual value function for each element that maps the principal's true utility to a virtual value. The agent with an element of largest virtual value is declared the winner, and their favorite outcome that still has greater virtual value than all other agents is accepted by the principal. We show that these mechanisms are dominant-strategy incentive compatible (\cref{thm:myerson-dsic}), and speculate whether they are optimal in an instance-by-instance sense, much like Myerson's revenue-optimal auction.

\subsection{Overview of Our Results}

As mentioned before, our work focuses on understanding the delegation gap, which is defined as the minimum over all instances of the ratio between the principal's optimal delegated expected utility and their optimal non-delegated expected utility. For the models we study in this paper, agents have no constraints on ``probing'' elements to learn their values, so the principal's optimal non-delegated utility is simply the maximum utility among all outcomes.

We start by showing that bounds on the delegation gap in the strategic case can be reduced to identical bounds in the adversarial case. This comes in two parts: the simple observation that delegating to strategic agents is at least as easy as delegating to adversarial agents (\cref{thm:adversarial-to-strategic-lower-bound}), and, less obviously, that for every adversarial instance within a central class, there is an analogous strategic instance with identical behavior from the principal's perspective (\cref{thm:adversarial-to-strategic-upper-bound}). This may be somewhat surprising, since it implies that any increase in utility from delegating to multiple agents is not, in general, attributable to strategic competition between those agents. Rather, the principal's utility seems to increase simply as a consequence of the larger pool of acceptable options afforded by a larger pool of agents.

Turning our focus toward adversarial delegation, we find a harsh $1/2$-approximation upper bound for any number of agents that carries over from the related single-agent model. This is due to the fact that the general form of the model allows for one agent to hold all elements that contribute non-zero expected utility to the principal, so, in essence, the principal must delegate to just that one agent.

However, moving beyond this impossibility, we show that when all agents have identical sets of elements, it is possible to achieve a competitive delegation gap of $1 - \O\left(\frac{\ln k}{k}\right)$. This is done in two parts: first achieving this approximation for instances with only atomless distributions (\cref{thm:atomless-adversarial-lower-bound}), and then showing how to modify the strategy to deal with atoms (\cref{thm:adversarial-lower-bound}). Notably, this approximation uses only a threshold mechanism, so it is simple to describe.

In the interest of demonstrating that other forms of symmetry also give competitive approximations, we show that the delegation gap of shuffled multi-agent delegation has the same $1 - \O\left(\frac{\ln k}{k}\right)$ lower-bound (\cref{thm:atomless-shuffled-adversarial-lower-bound} and \cref{thm:shuffled-adversarial-lower-bound}). Noting that a different symmetry assumption gives the same result, we conjecture that this is an instance of a more general phenomenon. 

Finally, we show that the optimal delegation gap achievable with $k$ agents in the agent-symmetric case is upper bounded by $1 - \Omega\left(\frac{1}{k}\right)$ (\cref{thm:adversarial-upper-bound}). We leave open for future work whether the gap between these upper and lower bounds can be closed.

\subsection{Related Work}

There is a relatively long history of delegation research in computer science and economics, notably starting over four decades ago with the work of Holmstrom \cite{holmstrom78,holmstrom80}. We refer readers to \cite{kleinberg2018delegated,bechtel2021probing,hajiaghayi2023agents} for a more detailed account of the followup work in delegation. A select sample includes the notable work of Alonso and Matouschek \cite{alonso2008optimal}, Armstrong and Vickers \cite{armstrong2010model}, Alonso et al. \cite{alonso2014resource}, and the recent work of Gan et al. \cite{gan2023optimal}. The last two study multi-agent delegation with two agents and no transfers, but they use different models and aim to find optimal mechanisms.

The past few years have seen a small resurgence in this area, initiated by the work of Kleinberg and Kleinberg \cite{KK18}. They study two models of delegation without payments, aiming for the same multiplicative approximation of the principal's optimal outcome, and show that the two problems can be reduced to known prophet inequalities and Pandora's box problems. This work was later expanded on by that of Bechtel and Dughmi \cite{bechtel2021probing}, who study a delegated model of stochastic probing with combinatorial constraints on the principal and agent, and Bechtel et al. \cite{bechtel2022pandora}, who explore different variants of the delegation of Pandora's box problems. This line of work has shown that delegation has close connections to prophet inequalities \cite{KS77,KS78,matroid_prophet,feldman2016online,prophet_easy,prophet_survey_lucier} and contention resolution schemes \cite{CRS,feldman2016online,ROCRS}, among other related problems.

Most similar to our work is that of Hajiaghayi et al. \cite{hajiaghayi2023agents}, who proposed (among other things) our main model of multi-agent delegation as a natural direction to build off of the existing work on delegation. Specifically, our strategic model is equivalent to their Bayesian mechanism multi-agent delegation with incomplete information. They show that when all agents have the same number of i.i.d. elements, the principal can achieve approximations tending to $1$ as $\alpha k m$ increases, where $k$ is the number of agents, $m$ is the number of elements per agent, and $\alpha$ is a parameter of the distributions. In contrast, we achieve an approximation tending to $1$ as $k$ increases when agents have symmetric sets of elements (not necessarily i.i.d.), with no conditions on the distributions or number of elements per agent. They also explore and achieve competitive approximations for different settings with varying levels of power for the principal and agents.

%%% Local Variables:
%%% mode: latex
%%% TeX-master: "main"
%%% End:

\section{Model}

Now we will formalize our three models of delegation in the following definitions.

\begin{definition}[Strategic Multi-Agent Delegation]
    \label{def:strategic}
    An instance of \emph{strategic multi-agent delegation} consists of $k$ agents labeled $1, \dots, k$, and associated with each agent $i$ are $m_i$ elements $e_{i,1}, \dots, e_{i,m_i}$. Each element $e$ has an ex-ante unknown independent random variable $V(e)$ sampled from distribution $\mu(e)$ over a set of \emph{outcomes} $\Omega(e)$ distinct from all other elements. Given an outcome $\omega \in \Omega(e)$, let $x(\omega)$ represent the principal's utility and $y(\omega)$ the utility of agent $i$ for outcome $\omega$.
    
    The principal starts by committing to a mechanism, which consists of sets of \emph{signals} $\Sigma_1, \dots, \Sigma_k$ for each of the agents, and a \emph{conditional acceptance function} $g : \Sigma_1 \times \dots \times \Sigma_k \to \Omega \union \{\bot\}$, where $\Omega$ is the set of all outcomes from all agents and $\bot$ represents the null or ``status quo'' outcome. Then each agent observes their own \emph{type}, the full set of random outcomes of all of their elements, and must select a signal to send to the principal. Finally, the principal uses $g$ to transform these signals into either a \emph{conditionally accepted} outcome $\omega \in \Omega$ or a rejection of all outcomes $\bot$. If they conditionally accept outcome $\omega$ and this outcome was observed by agent $i$, then they receive utility $x(\omega)$, agent $i$ receives utility $y(\omega)$, and all other agents receive utility $0$. If they reject all or conditionally accept an unobserved outcome, then all players receive utility $0$.
    
    All players act to maximize their expected utility, breaking ties in favor of the principal, and have common knowledge of the setup including the distributions of all elements but excluding the realizations of random variables. We also assume that agents can communicate only with the principal and cannot collude with each other. We require that all random variables $V(e)$ are mutually independent, but the principal's and agent's values for a single element can be arbitrarily dependent.
\end{definition}

Given this setup, we will define some additional terminology and notation. If the principal (conditionally) accepts an outcome from an agent, then that agent is called the \emph{(conditional) winner}. All other agents are called \emph{losers}.\footnote{We apologize in advance to all such agents.} Let $\mathcal{T}$ be the set of all possible types for all agents, with $\mathcal{T}_i$ being the set of types for agent $i$. Also let $\Omega$ be the set of all outcomes, $\Omega_i$ the set of outcomes from agent $i$, $E$ the set of all elements, and $E_i$ the set of elements of agent $i$.

Given a particular instance $I$, mechanism $M$, and strategies $\sigma$ for all agents, let the \emph{allocation} function $f_{I,M,\sigma}(t)$ yield the outcome accepted by the principal when agents have full type $t \in \mathcal{T}$. Furthermore, let $\princ_{I,M,\sigma}$ represent the random utility of the delegating principal and $\opt_I$ represent the random utility of the non-delegating principal, i.e. $\opt = \max_{e \in E} x(V(e))$. We may omit subscripts when the meaning is clear from context. Then we can write the delegation gap as
\begin{equation*}
    \min_I \max_{M, \sigma} \frac{\E[\princ_{I,M,\sigma}]}{\E[\opt_I]},
\end{equation*}
where $I$, $M$, $\sigma$ range over all instances, mechanisms, and equilibria, respectively, that are in consideration. Observe that we define the delegation gap in terms of the best equilibrium for the principal, as is often done in mechanism design.

Note that the principal's acceptance of an outcome is contingent on learning its true value, i.e. agents cannot lie about outcomes they want the principal to accept. This assumption follows previous work on such problems, is necessary to reign in worst-case agents, and is natural for many applications of this problem. Specifically, proposed solutions to a problem are often verifiable to the principal, even if it is costly to do so. For example, in our grant proposal scenario, the committee can form a panel of experts to verify the claims of a favored proposal.

In this paper, we usually consider a restricted model in which all agents have equivalent sets of elements, i.e. $m_1 = m_2 = \dots = m_k = m$ and $V(e_{i,j})$ is identically distributed to $V(e_{i',j})$ for agents $i \ne i'$. We call this \emph{agent-symmetric} multi-agent delegation. Sometimes we will consider the even more restricted model in which all elements are identically distributed, i.e. $V(e_{i,j})$ is identically distributed to $V(e_{i',j'})$ for agents $i, i'$ and elements $j, j'$. This is called \emph{fully-symmetric} multi-agent delegation.

We now consider a version of the same model in which agents adversarially aim to minimize the principal's utility.

\begin{definition}[Adversarial Multi-Agent Delegation]
    \label{def:adversarial}
    \emph{Adversarial multi-agent delegation} is the same as strategic multi-agent delegation, except that agents' utilities (in the form of $y$-values) are left undefined. Instead of maximizing their own expected utility, agents aim to minimize the principal's expected utility subject to maintaining a positive probability of winning.
\end{definition}

Observe that a ``truly'' adversarial group of agents would have no incentive to share any information with the principal, thereby guaranteeing everyone $0$ utility. This is unhelpful for our purposes, so we make the weak assumption that agents will only take actions for which there remains some possibility of them winning.

Finally, we will define a variant of each of the two models above that ``shuffles'' each agent's elements before the game starts.

\begin{definition}[Shuffled Strategic / Adversarial Multi-Agent Delegation]
    \label{def:shuffled}
    Given either the strategic or adversarial model, we can also consider a \emph{shuffled} version of the same model. The model is the same except that there is a pool of $n$ elements, each of which is given to a uniformly random agent independently of other elements. Each agent knows only which elements they received, and the principal knows nothing about who received what. Everything proceeds as before, but we measure the principal's expected utility by taking an expectation over the random distribution of elements.
\end{definition}

Observe that any mechanism for an instance of shuffled multi-agent delegation must be oblivious to which agent holds which elements, since that information is not available to the principal when they declare the mechanism.

\subsection{Mechanisms}

While some previous work has explored randomized mechanisms, in this work we consider only deterministic mechanisms. We start with a class of ``single-proposal'' mechanisms where each agent proposes only a single outcome to the principal, and then the principal picks a winner from among those outcomes or otherwise rejects them all. This builds on the single-agent single-proposal mechanism introduced by \cite{kleinberg2018delegated} and is identical to the multi-agent version proposed in \cite{hajiaghayi2023agents}.

\begin{definition}[Single-Proposal Mechanism]
    \label{def:single-proposal}
    Fix any instance of multi-agent delegation with $k$ agents. Then a \emph{single-proposal mechanism} for this instance consists of a set $R_i \subseteq \Omega_i$ for each agent $i$, along with a deterministic tie-breaking order $\prec$ over $E$. The principal commits to this mechanism, and each agent $i$ responds by proposing a single element $e$ along with its observed outcome $\omega \in \Omega(e)$. The principal will accept the outcome $\omega$ from some agent $i$ for which $\omega \in R_i$ and $x(\omega)$ is maximized, using the tie-breaking order $\prec$ if there are multiple such outcomes. If no such outcome exists, then the principal rejects.
\end{definition}

We note that the tie-breaking order is a necessary part of the definition to ensure that the principal's mechanism is deterministic and consistent, but it ultimately does not affect the principal's utility, so we will usually omit defining it for single-proposal mechanisms.

While it may seem overly restrictive to only allow a single proposal from each agent, it turns out that this is a powerful class of mechanisms capable of performing at least as well as any other deterministic mechanism as originally defined in \cref{def:strategic}. This is formalized in the following proposition of \cite{hajiaghayi2023agents}, who proved it in the multi-agent case.

\begin{lemma}[Theorem 4.11 of \cite{hajiaghayi2023agents}]
    \label{thm:single-proposal}
    For any mechanism $M$ with an arbitrary equilibrium $\sigma$, there exists a single-proposal mechanism $M'$ such that for any equilibrium $\sigma'$, full type $t \in \mathcal{T}$, and accepted outcomes $\omega = f_{M,\sigma}(t)$ and $\omega' = f_{M',\sigma'}(t)$, we have $x(\omega) \le x(\omega')$.
\end{lemma}

One clear drawback with single-proposal mechanisms is that competitive strategies may consist of complex acceptable sets of outcomes, i.e. $R_i$ sets, and this complexity can make implementation difficult. A special case of single-proposal mechanisms that avoids this complexity is the class of threshold mechanisms.

\begin{definition}[Threshold Mechanism]
    \label{def:threshold}
    A \emph{threshold mechanism} is a special case of single-proposal mechanism in which each set $R_i$ is defined by a mapping $\theta : E \to \R$ from each agent's elements to thresholds such that an outcome $\omega$ from element $e$ is in $R_i$ if and only if $x(\omega) \ge \theta(e)$ (we also allow strict thresholds $x(\omega) > \theta(e)$).
\end{definition}

If a threshold mechanism is such that $\theta(e_{i,j}) = \theta(e_{i,j'})$ for all agents $i$ and all $j \ne j'$, then we call the mechanism \emph{element-symmetric}. Similarly, if it satisfies $\theta(e_{i,j}) = \theta(e_{i',j})$ for all agents $i \ne i'$ and all $j$, then we call it \emph{agent-symmetric}. A mechanism that is both agent-symmetric and element-symmetric is called \emph{fully-symmetric}. We may emphasize when neither property necessarily holds by calling a threshold mechanism \emph{general}.

Lastly, for completeness, we will also define a natural direct revelation mechanism inspired by Myerson mechanisms in auction design.

\begin{definition}[Myerson-type Mechanism]
    \label{def:myerson-type}
    Fix an instance of multi-agent delegation with $k$ agents. Then a \emph{Myerson-type mechanism} for this instance is a direct revelation mechanism consisting of a \emph{virtual value} function $\phi(e) : \R_{\ge 0} \to \R_{\ge 0}$ for every element $e$, and a tie-breaking total order $\prec$ over all elements $E$.
    
    The principal commits to this mechanism, and each agent $i$ responds by revealing a (possibly false) type $\hat{\omega}_{i,1}, \dots, \hat{\omega}_{i,m_i}$. For any element $e_{i,j}$, let $v_{i,j} = \phi_{i,j}(x(\hat{\omega}_{i,j}))$ be its virtual value according to that reported type, and for any agent $i$, let $v_i = \max_j v_{i,j}$ be the best virtual value among their elements. Let $v^* = \max_i v_i$ be the maximum such value among all agents and $i^*$ be an agent with maximum virtual value $v_{i^*} = v^*$.
    \begin{enumerate}
        \item If $v^* < 0$, then the principal rejects.
        \item If $v^* \ge 0$ and $i^*$ is not unique, then the principal conditionally accepts the outcome $\omega_{i,j}$ for which $v_{i,j} = v^*$ that appears earliest in the tie-breaking order.
        \item If $v^* \ge 0$ and $i^*$ is unique, then let $j^*$ be such that $y_{i^*,j^*}$ is maximized subject to $v_{i^*,j^*} \ge 0$ and $v_{i^*,j^*} > v_i$ for all agents $i \ne i^*$. The principal conditionally accepts an arbitrary such outcome $\omega_{i^*,j^*}$.
    \end{enumerate}
\end{definition}

Note that the behavior of the mechanism can be described simply as: pick the agent with largest virtual value as the winner, and conditionally accept their favorite outcome that has larger virtual value than any other agent. Unfortunately, we have to be careful about how exactly to handle ties to ensure that the mechanism remains deterministic and truthful. We note that just as in the case with single-proposal mechanisms, the tie-breaking order has no effect on the principal's expected utility.

We claim that agents under this mechanism are incentivized to report their true types.

\begin{proposition}[Myerson-type Mechanisms are Truthful]
    \label{thm:myerson-dsic}
    Myerson-type mechanisms are dominant strategy incentive compatible.
\end{proposition}

\begin{proof}
    Fix some agent $i$ for which we will show that a dominant strategy is to report their true type. For all other agents $i'$, let $\sigma_{i'} : \mathcal{T}_{i'} \to \mathcal{T}_{i'}$ be their strategy, i.e. a mapping from their true type to their reported type. Given these strategies, let $\bar{y}(t, t')$ be expected utility of agent $i$ when their true type is $t$ and they report type $t'$. This expectation is taken over randomness in the other agent's types, strategies, and the mechanism itself. We aim to show that $\bar{y}(t, t) \ge \bar{y}(t, t')$ for all $t, t' \in \mathcal{T}_i$. Let $t = (\omega_{i,1}, \dots, \omega_{i,m_i})$ and $t' = (\omega'_{i,1}, \dots, \omega'_{i,m_i})$. For simplicity, we will write $y_{i,j} = y(\omega_{i,j})$ and $y'_{i,j} = y(\omega'_{i,j})$, and analogously for $x$-values.
    
    First, we will consider the case when the agent lies about outcome $\omega'_{i,j}$ such that $y'_{i,j} \ne y_{i,j}$ for some index $j$. Observe that the value of $y'_{i,j}$ does not affect whether or not agent $i$ will be picked as the winner. It can only affect which outcome is accepted by the principal in case 3 of the mechanism. Either outcome $\omega_{i,j}$ is conditionally accepted and the principal detects the lie, $\omega_{i,j}$ is not accepted and an outcome $\omega_{i,j'}$ with $y'_{i,j} \le y'_{i,j'} \le y_{i,j}$ was conditionally accepted instead, or $e_{i,j}$ is not accepted and the value of $y'_{i,j}$ has no effect on the accepted outcome. In all three cases, agent $i$ does at least as well by reporting an outcome with the true value of $y_{i,j}$. Therefore, we can assume that $y'_{i,j} = y_{i,j}$ and show that $\bar{y}(t, t) \ge \bar{y}(t, t')$ in this case.
    
    Now consider when the agent lies about an outcome $\omega'_{i,j}$ such that $x'_{i,j} \ne x_{i,j}$ for some index $j$. Let $\tau$ be the maximum reported virtual value among all agents other than agent $i$. We will consider four cases.
    \begin{enumerate}
        \item The principal conditionally accepts outcome $\omega_{i,j}$, they detect the lie, and agent $i$ receives $0$ utility.
        \item The principal does not accept any outcome from agent $i$, and the agent receives $0$ utility.
        \item The principal conditionally accepts a distinct outcome $\omega_{i,j'}$ and $\phi_{i,j}(x'_{i,j}) < \tau$.
        \item The principal conditionally accepts a distinct outcome $\omega_{i,j'}$ and $\phi_{i,j}(x'_{i,j}) \ge \tau$. Then $y_{i,j'} \ge y_{i,j}$.
    \end{enumerate}
    It is easy to check that in all of the above cases, agent $i$ again does at least as well by reporting an outcome with the true of $x_{i,j}$. Therefore, they do at least as well by reporting the true outcome $\omega_{i,j}$. This holds regardless of the strategies of other agents, so Myerson-type mechanisms are dominant-strategy incentive-compatible.
\end{proof}

%%% Local Variables:
%%% mode: latex
%%% TeX-master: "main"
%%% End:

\section{Strategic Agents}

It turns out that it is possible to understand the delegation gap of large classes of strategic multi-agent delegation without considering strategic behavior and equilibria. This is by first reducing it to the delegation gap of adversarial multi-agent delegation, and then studying the adversarial model.

\begin{definition}
    \label{def:analogous}
    Given an instance $I$ of strategic multi-agent delegation and an instance $I'$ of adversarial multi-agent delegation, we say that $I$ and $I'$ are \emph{analogous} if they are equivalent, up to isomorphism, aside from the $y$-values (which $I'$ does not define).
\end{definition}

Observe that for a particular instance of the strategic model, there's only one (up to isomorphism) analogous instance of the adversarial model. Now we can state the following rather intuitive result that allows us to translate lower-bounds on the adversarial delegation gap to lower-bounds on the strategic delegation gap:

\begin{lemma}[Adversarial to Strategic Lower Bound]
    \label{thm:adversarial-to-strategic-lower-bound}
    For every instance $I$ of strategic multi-agent delegation, the analogous instance $I'$ of adversarial multi-agent delegation is such that for every single-proposal mechanism $M'$ for $I'$, there exists a mechanism $M$ and induced equilibrium $\sigma$ for $I$ for which
    \begin{equation*}
        x(f_{I,M,\sigma}(t)) \ge x(f_{I',M'}(t))
    \end{equation*}
    for all sets of types $t$.
\end{lemma}

\begin{proof}
    Take an arbitrary such strategic instance $I$, the analogous adversarial instance $I'$, and any single-proposal mechanism $M'$ for $I'$. Let $M = M'$ and consider any equilibrium $\sigma$ of $M$ on instance $I$. Finally, fix an arbitrary set of types $t$.
    
    Consider a particular observed outcome $\omega_i$ from agent $i$. We claim that if $\omega_i$ has $0$ probability of winning for the adversarial agent from their perspective, then it also has $0$ probability of winning for the strategic agent. This is because either (1) $\omega_i$ is unacceptable to the principal under mechanism $M'$ or (2) there is no set of types for other agents in which $\omega_i$ would win. In either case, this means that the strategic agent also has $0$ probability of winning under mechanism $M$ and equilibrium $\sigma$.
    
    Therefore, we can see that any outcome that the strategic agent might propose is also considered by the adversarial agent. And since the adversarial agent always picks the principal's least favorite, the desired inequality follows.
\end{proof}

Conversely, we want to be able to upper-bound the strategic delegation gap by the adversarial delegation gap, which we can do even in the fully-symmetric setting. Since our models include the fully-symmetric setting as a special case and our upper bound on the adversarial setting in \cref{thm:adversarial-upper-bound} is fully-symmetric, the following result shows that this same bound carries over to the strategic setting, yielding \cref{thm:strategic-upper-bound}.

\begin{lemma}[Adversarial to Strategic Upper Bound]
    \label{thm:adversarial-to-strategic-upper-bound}
    For every instance $I$ of fully-symmetric adversarial multi-agent delegation with finite-support distributions, there exists an analogous instance $I'$ of strategic delegation such that for every single-proposal mechanism $M'$ and induced equilibrium $\sigma'$ for $I'$, there exists a single-proposal mechanism $M$ for $I$ for which
    \begin{equation*}
        f_{I,M} = f_{I',M',\sigma'}.
    \end{equation*}
\end{lemma}

\begin{proof}
    Take an arbitrary instance $I$ of fully-symmetric adversarial multi-agent delegation with finite-support distributions. We will describe how to construct $I'$ by defining the agent's utility for each possible outcome. Then we will show that any single-proposal mechanism $M'$ on that instance induces a unique equilibrium under which agents act adversarially.
    
    Fix any agent $a$ and the set of possible outcomes $\Omega_a$ that they might observe. Since all (identical) distributions have finite support, $\Omega_a$ is finite. Label all outcomes in $\Omega_a$ by $\omega_1, \dots, \omega_\ell$ such that $x(\omega_1) \le \dots \le x(\omega_\ell)$. Let $P$ to be the probability that all other agents observe only equivalent copies of outcome $\omega_1$, and define agent $i$'s utilities such that $y(\omega_i) = (P / 2)^i$ for all $1 \le i \le \ell$. This defines the analogous instance $I'$. We note that $P > 0$, so this definition guarantees that $y(\omega_1) > \dots > y(\omega_\ell)$.
    
    Now take an arbitrary single-proposal mechanism $M'$ and induced equilibrium $\sigma'$ for strategic instance $I'$. We will choose $M$ to be an identical mechanism for adversarial instance $I$, and we now aim to show that strategic agents under mechanism $M'$ and equilibrium $\sigma'$ act identically to adversarial agents under mechanism $M$.
    
    Let $Q(\omega_i)$ be the probability that proposal $\omega_i$ from agent $a$ wins under mechanism $M'$ and equilibrium $\sigma'$. We claim that $Q(\omega_i) \ge P$ whenever $Q(\omega_i) > 0$. This is because the mechanism is deterministic, so if there is a positive probability that proposal $\omega_i$ wins, it must occur at least when all other agents observe only the principal's least favorite outcome $\omega_1$.
    
    Now consider two different outcomes $\omega_i$ and $\omega_j$ where $i < j$ and both outcomes have a positive probability of being accepted by the principal under mechanism $M'$ and equilibrium $\sigma'$, i.e. $Q(\omega_i) > 0$ and $Q(\omega_j) > 0$. By the ordering of elements, we have $x(\omega_i) \le x(\omega_j)$ and $y(\omega_i) > y(\omega_j)$. A strategic agent will prefer proposing $\omega_i$ over $\omega_j$ if doing so will strictly increase their expected utility:
    \begin{align*}
        y(\omega_j) \cdot Q(\omega_j)
        &= (P / 2)^j \cdot Q(\omega_j) \\
        &= y(\omega_i) \cdot (P / 2)^{j - i} \cdot Q(\omega_j) \\
        &< y(\omega_i) \cdot P \\
        &\le y(\omega_i) \cdot Q(\omega_i),
    \end{align*}
    where the last inequality holds because $Q(\omega_i) > 0$ (so $Q(\omega_i) \ge P$) and the second to last because $Q(\omega_j) > 0$, $0 < P \le 1$, and $j > i$.
    
    Therefore, agent $i$ will always strictly prefer proposing outcome $\omega_i$ over $\omega_j$ when they have the opportunity to do so. Recall that $x(\omega_i) \le x(\omega_j)$, so this means that agent $i$ will always propose the principal's least favorite outcome which has a positive probability of being accepted. This is precisely what it means to be an adversarial agent, and since this strategy is optimal regardless of the behavior of other agents, it is a dominant strategy. All agents have the same dominant strategy, so there is a unique equilibrium under which all agents act adversarially with respect to mechanism $M$. Therefore, the selection functions $f_{I,M}$ and $f_{I',M',\sigma'}$ are identical.
\end{proof}

We claim that the same result can be proven in the case of atomless distributions with bounded probability density, though the details of the proof are somewhat different. Finally, this result can likely be extended to even larger classes of instances, but it is sufficient for our purposes as stated.

%%% Local Variables:
%%% mode: latex
%%% TeX-master: "main"
%%% End:

\section{Adversarial Agents}

As observed by Hajiaghayi et al., the general model of multi-agent delegation we presented has a harsh delegation gap upper bound of $1/2$.

\begin{remark}
    \label{thm:asymmetric-adversarial-upper-bound}
    For any $\varepsilon > 0$, there exists an instance of multi-agent delegation for which the principal cannot achieve better than a $1/2 + \varepsilon$ approximation.
\end{remark}

We omit the proof, but this result comes from a known upper bound in the single-agent case which can be easily adapted to the multi-agent case since there are no restrictions on how elements and the principal's utility are spread among agents. Intuitively, this model allows arbitrary asymmetry between agents, so that all of the principal's optimal expected utility can come from a single agent.

This impossibility motivates our exploration of the delegation gap in symmetric multi-agent delegation. We will first show that we can achieve a competitive approximation for agent-symmetric adversarial multi-agent delegation. The proof method is fairly straightforward and very similar to proofs of classic prophet inequalities. Note that all lower-bounds in this section are $p$-approximations, where $p$ is the unique positive real solution to $p^k + p - 1 = 0$. It is not too difficult to show that $p = 1 - \Theta(\frac{\ln k}{k})$, so all our mechanisms achieve this approximation factor. We omit the proof of this simple fact due to space constraints, and relegate it to the full archival version of the paper.

\begin{proposition}[Atomless Adversarial Lower Bound]
    \label{thm:atomless-adversarial-lower-bound}
    Let $p$ be the unique positive real solution to $p^k + p - 1 = 0$ for an arbitrary $k > 0$. There exists a $p$-approximate threshold mechanism for adversarial agent-symmetric multi-agent delegation with $k$ agents and atomless distributions.
\end{proposition}

\begin{proof}
    Let $X^{\max}_i = \max_{e \in E_i} x(V(e))$ be the (random) maximum value among agent $i$'s elements, and let $X^* = \max_i X^{\max}_i$ be the maximum value among all elements. Since this instance is agent-symmetric, $V(e_{i,j})$ is identically distributed to $V(e_{i',j})$, and therefore $X^{\max}_i$ is identically distributed to $X^{\max}_{i'}$.
    
    Choose an arbitrary $p \in [0, 1]$, which we will fix later, and let $t$ be a corresponding threshold such that $\Pr[X^{\max}_i < t] = p$. Consider what happens when we run a fully-symmetric threshold mechanism with threshold $t$.
    
    The probability that at least one element meets threshold $t$ is $1 - p^k$. In that case, we are guaranteed that some agent will propose a value of at least $t$, so we are guaranteed utility at least $t$. Now consider an element of value $X^*$ and the agent holding it. With probability at least $p$, all of the agent's other elements do not meet the threshold, and this is true regardless of the value of $X^*$. If $X^*$ itself does not meet the threshold then we get $0$ utility, but if it does, then we get an additional $X^* - t$ over the guaranteed threshold. Therefore,
    \begin{align*}
        \E[\princ]
        &\ge (1 - p^k) t + p \E[(X^* - t)^+] \\
        &\ge (1 - p^k) t + p (\E[X^*] - t)^+.
    \end{align*}
    
    Now, if we choose $p \in [0, 1]$ such that $p = 1 - p^k$, i.e. $p^k + p - 1 = 0$, then we get
    \begin{equation*}
        \E[\princ] \ge p \E[X^*] = p \E[\opt].
    \end{equation*}
    
    Since there is a unique solution to $p^k + p - 1 = 0$ for which $0 \le p \le 1$, we will choose $p$ to be this value. Then the mechanism that sets the corresponding threshold $t$ for all elements achieves a $p$-approximation of the optimal non-delegated utility.
\end{proof}

Observe that this is a fully-symmetric threshold mechanism, setting the same threshold for all elements. Notice also that the mechanism as described only works when there exists a threshold $t$ such that $\Pr[X^{\max}_i < t] = p$. This is guaranteed for atomless distributions, but not in general. Nevertheless we can extend this result to include distributions with atoms.

It might be tempting to guess that setting distinct thresholds for different agents could afford strictly more power and achieve a better approximation. However, this does not appear to be true, and we found that a more involved analysis for the general case achieves no better than $1 - \Theta(\ln k / k)$.

The above proof breaks for certain distributions with atoms, but we will now fix that, obtaining the following result:
\begin{theorem}[Adversarial Lower Bound]
    \label{thm:adversarial-lower-bound}
    Let $p$ be the unique positive real solution to $p^k + p - 1 = 0$ for an arbitrary $k > 0$. There exists a $p$-approximate threshold mechanism for adversarial agent-symmetric multi-agent delegation with $k$ agents.
\end{theorem}

\begin{proof}
    Fix an arbitrary agent $i$, and recall that we want to set a threshold $t$ such that $\Pr[X^{\max}_i < t] = p$. We will assume that this is not possible, i.e. there exists a value $x'$ such that $\Pr[X^{\max}_i < x'] < p \le \Pr[X^{\max}_i \le x']$. Define
    \begin{equation*}
        \varphi = \frac{p - \Pr[X^{\max}_i < x']}{\Pr[X^{\max}_i \le x'] - \Pr[X^{\max}_i < x']} = \frac{p - \Pr[X^{\max}_i < x']}{\Pr[X^{\max}_i = x']},
    \end{equation*}
    where $\Pr[X^{\max}_i = x'] > 0$ by our assumption. Observe that $\varphi > 0$ since $p > \Pr[X^{\max}_i < x']$, and that $\varphi \le 1$ since $p \le \Pr[X^{\max}_i \le x']$. This is important because we will later treat $\varphi$ as a probability.
    
    Imagine modifying the distribution by splitting the probability of outcome $x'$ into two distinguishable outcomes $x^-$ and $x^+$ with the same value as $x'$. Let their probabilities be such that $\Pr[X^{\max}_i = x^-] = \varphi \Pr[X^{\max}_i = x']$ and $\Pr[X^{\max}_i = x^+] = (1 - \varphi) \Pr[X^{\max}_i = x']$. If we could set a threshold $t$ such that $x^- < t < x^+$, then
    \begin{align*}
        \Pr[X^{\max}_i < t]
        &= \Pr[X^{\max}_i < x^-] + \Pr[X^{\max}_i = x^-] \\
        &= \Pr[X^{\max}_i < x^-] + \varphi \Pr[X^{\max}_i = x'] \\
        &= \Pr[X^{\max}_i < x^-] + (p - \Pr[X^{\max}_i < x^-]) \\
        &= p.
    \end{align*}
    In this situation, our earlier analysis still works and we achieve the desired approximation. To complete the proof, we will describe a mechanism that simulates this situation, and then show how to implement it with a threshold mechanism.
    
    Consider the following mechanism: the principal sets a threshold of $t = x'$ and a randomization probability of $\varphi$. Outcomes equal to the threshold $t$ are considered unacceptable to each agent independently with probability $\varphi$, and all other outcomes are considered acceptable if and only if they meet threshold $t$. To determine if an outcome is acceptable or not, each agent will flip a biased coin and must report the result honestly.
    
    Now, by the principle of deferred decisions, the mechanism in which each agent flips a biased coin is equivalent to a mechanism in which the principal flips each agent's coin in advance and notifies them of the result as part of the mechanism. Interpret this randomized mechanism as a randomization over deterministic mechanisms with the same approximation factor. Since delegation is a Stackelberg game, a randomized mechanism that reveals its coins is no better than the best deterministic mechanism it randomizes over. Therefore, there exists some deterministic threshold mechanism achieving approximation $p$.
\end{proof}

Observe that the resulting mechanism may set different thresholds for different agents (where each one is either a strict threshold $>t$ or a weak threshold $\ge t$), so it is not agent-symmetric in general. Additionally, the proof is not fully constructive and does not tell us which threshold to set for which agent. However, by observing that the instance is agent-symmetric, we can see that the principal's approximation only depends on the number of agents that are given each type of threshold. Therefore, the principal only has to manually analyze $k + 1$ different mechanisms to find the best choice.

Applying \cref{thm:adversarial-to-strategic-lower-bound} gives us the following corollary:
\begin{corollary}[Strategic Lower Bound]
    Let $p$ be the unique positive real solution to $p^k + p - 1 = 0$ for an arbitrary $k > 0$. There exists a $p$-approximate threshold mechanism for strategic agent-symmetric multi-agent delegation with $k$ agents.
\end{corollary}

Having achieved a competitive approximation for the agent-symmetric setting, we consider a different way to get around the harsh $\frac{1}{2}$-approximation upper-bound on general multi-agent delegation. Intuitively, this limitation occurs because agents' elements can be arbitrarily asymmetric, with some agents contributing far more to the principal's utility than others. One natural way of getting around this is to randomize which agents get which elements. This motivates our shuffled model and the following result:

\begin{proposition}[Atomless Shuffled Adversarial Lower Bound]
    \label{thm:atomless-shuffled-adversarial-lower-bound}
    Let $p$ be the unique positive real solution to $p^k + p - 1 = 0$ for an arbitrary integer $k > 0$. There exists a $p$-approximate threshold mechanism for adversarial multi-agent delegation with $k$ agents and $n$ (not necessarily i.i.d.) atomless elements assigned uniformly at random to the agents.
\end{proposition}

\begin{proof}
    The proof is similar to that of \cref{thm:atomless-adversarial-lower-bound}. Choose an arbitrary $q \in [0, 1]$, which we will fix later, and let $t$ be a threshold such that $\Pr[X^{\max} < t] = q$. We will also define $q_e = \Pr[X_e < t]$ such that $q = \prod_{e \in E} q_e$. Consider running a fully symmetric threshold mechanism with threshold $t$.
    
    The probability that at least one element meets the threshold is $1 - q$. When that happens, we are guaranteed a utility of at least $t$. Now consider an element of value $X^*$ and the agent $i$ holding it. Let $E_i$ be the random set of elements that agent $i$ receives, and consider the probability that none of their elements meets the threshold. We have
    \begin{align*}
        \Pr[X^{\max}_i < t]
        &= \prod_{e \in E} \big( \Pr[e \in E_i] \Pr[X_e < t] + \Pr[e \notin E_i] \big) \\
        &= \prod_{e \in E} \left( \frac{1}{k} q_e + 1 - \frac{1}{k} \right) \\
        &= \prod_{e \in E} \left( 1 + (q_e - 1) \frac{1}{k} \right) \\
        &\ge \prod_{e \in E} q_e^{1/k} \\
        &= q^{1/k},
    \end{align*}
    where the second to last step follows from Bernoulli's inequality.
    
    Following the same kind of analysis as before, the principal's utility is
    \begin{align*}
        \E[\princ]
        &\ge (1 - q) t + q^{1/k} \E[(X^* - t)^+] \\
        &\ge (1 - q) t + q^{1/k} (\E[X^*] - t)^+.
    \end{align*}
    
    Let $p = 1 - q$. By choosing $q \in [0, 1]$ such that $1 - q = q^{1/k}$, i.e. $p^k + p - 1 = 0$, we get the desired approximation $\E[\princ] \ge p \E[\opt]$.
\end{proof}

We omit the details but observe that the same approximation holds when a slightly different form of randomization is used: each agent is given a uniformly random set of exactly $n / k$ elements. This seems to imply that there exists a more general randomized model for which we can achieve the same approximation lower bound.

Note also that the same method of dealing with atoms in distributions works for the shuffled model. This implies the following result:

\begin{theorem}[Shuffled Adversarial Lower Bound]
    \label{thm:shuffled-adversarial-lower-bound}
    Let $p$ be the unique positive real solution to $p^k + p - 1 = 0$ for an arbitrary $k > 0$. There exists a $p$-approximate threshold mechanism for adversarial multi-agent delegation with $k$ agents and $n$ (not necessarily i.i.d.) elements assigned uniformly at random to the agents.
\end{theorem}

We complement our results lower-bounding the delegation gap in the symmetric case with a new upper bound. Given the gap between these bounds, at least one is not tight, and we leave open for future work where exactly the delegation gap of this model lies.

\begin{theorem}[Adversarial Upper Bound]
    \label{thm:adversarial-upper-bound}
    There does not exist a multi-agent delegation mechanism achieving strictly better than a $\left(1 - \frac{1}{2k + 1}\right)$-approximation for arbitrary instances of fully-symmetric adversarial multi-agent delegation with $k$ agents.
\end{theorem}

\begin{proof}
    Fix an arbitrary number of agents $k \ge 2$, and consider a small $0 < \varepsilon < 1$. We will construct an instance for which the principal cannot achieve better than $\left(1 - \frac{1 - \O(\varepsilon)}{2k + 1}\right)$. Then, by taking the limit $\varepsilon \to 0$, we get the desired result.
    
    For every agent $i$, let them have two elements $e_{i,1}$ and $e_{i,2}$ distributed i.i.d.\ with two outcomes $\omega_L$ and $\omega_H$ such that $\Pr[V(e_{i,j}) = \omega_L] = 1 - \varepsilon$ and $\Pr[V(e_{i,j}) = \omega_H] = \varepsilon$. Let $x(\omega_L) = 1$ and $x(\omega_H) = 1 / \varepsilon$, and we will call these outcomes \emph{low} and \emph{high}, respectively.
    
    Now consider the principal's optimal non-delegated utility on this instance. If any outcome is high, the principal should take one such outcome. Otherwise, at least one outcome will be low. It is straightforward to show that this gives the principal an expected utility of
    \begin{align*}
        \E[\opt]
        &= \frac{1}{\varepsilon} \left( 1 - (1 - \varepsilon)^{2k} \right) + (1 - \varepsilon)^{2k} \\
        &= \frac{1}{\varepsilon} (2k\varepsilon + \O(\varepsilon^2)) + 1 + \O(\varepsilon) \\
        &= 2k + 1 + \O(\varepsilon).
    \end{align*}
    
    It is straightforward but somewhat tedious to understand the principal's optimal delegation mechanism. Recall that by \cref{thm:single-proposal} we only have to consider single-proposal mechanisms, and such a mechanism declares an acceptable set of outcomes for each agent. For any particular element, there are two possible outcomes: $\omega_H$ and $\omega_L$. So the acceptable set of outcomes for any particular element must be one of: $\emptyset$, $\{\omega_H\}$, $\{\omega_L\}$, and $\{\omega_H, \omega_L\}$. It is not hard to see that, from the principal's perspective, $\emptyset$ is Pareto-dominated by $\{\omega_H\}$ and $\{\omega_L\}$ is Pareto-dominated by $\{\omega_H, \omega_L\}$. Therefore, it is without loss of generality that we can consider mechanisms that choose either $H = \{\omega_H\}$ (representing a high threshold) or $L = \{\omega_H, \omega_L\}$ (representing a low threshold) as the acceptable outcomes for a particular element.
    
    Now, consider the two elements held by any agent. The principal can accept any of the following combinations of outcomes: $L \times L$, $L \times H$ (which is equivalent to $H \times L$ by symmetry), and $H \times H$. Here we observe that $L \times L$ is Pareto-dominated by $L \times H$, so it is without loss to consider only $L \times H$ and $H \times H$.
    
    Finally, consider any two agents. We can think of acceptable set $L \times H$ as trading off the probability of receiving a high value for the guarantee that we at least receive the low value. Therefore, we can see that choosing acceptable sets $(L \times H, L \times H)$ for two agents is Pareto-dominated by $(L \times H, H \times H)$. In other words, we need only consider two mechanisms: ($A$) accepting $H \times H$ from all agents or ($B$) accepting $L \times H$ from a single agent and $H \times H$ from all others.
    
    For mechanisms $A$ and $B$, the principal receives an expected utility of, respectively,
    \begin{align*}
        \E[\princ_A] &= \frac{1}{\varepsilon} (1 - (1 - \varepsilon)^{2k}), \\
        \E[\princ_B] &= \frac{1}{\varepsilon} (1 - (1 - \varepsilon)^{2k - 1}) + (1 - \varepsilon)^{2k - 1}.
    \end{align*}
    Both cases simplify to
    \begin{equation*}
        \E[\princ] = 2k + \O(\varepsilon).
    \end{equation*}
    Therefore, the delegation gap for this instance is
    \begin{equation*}
        \frac{2k + \O(\varepsilon)}{2k + 1 + \O(\varepsilon)} = 1 - \frac{1 - \O(\varepsilon)}{2k + 1},
    \end{equation*}
    completing the proof.
\end{proof}

Finally, applying \cref{thm:adversarial-to-strategic-upper-bound} gives us the following corollary:
\begin{corollary}[Strategic Upper Bound]
    \label{thm:strategic-upper-bound}
    There does not exist a multi-agent delegation mechanism achieving strictly better than a $\left(1 - \frac{1}{2k + 1}\right)$-approximation for arbitrary instances of fully-symmetric strategic multi-agent delegation with $k$ agents.
\end{corollary}

%%% Local Variables:
%%% mode: latex
%%% TeX-master: "main"
%%% End:

\section{Conclusion and Open Questions}

We conclude by listing some open questions and directions for future work:
\begin{enumerate}
    \item Which, if either, of our bounds on the delegation gap are tight?
    
    \item Can we get a full characterization of optimal mechanisms for strategic multi-agent delegation? Could Myerson-type mechanisms be optimal?
    
    \item How do these results change when considering the larger class of randomized mechanisms? Does the delegation gap improve?
    
    \item Previous work \cite{kleinberg2018delegated,bechtel2021probing,bechtel2022pandora} has considered single-agent delegation in the presence of ``probing constraints'' such as probing costs and combinatorial constraints on sets of probed elements. How do these affect multi-agent delegation?
    
    \item Previous work has also considered single-agent delegation in which the principal can accept sets of outcomes subject to a hard constraint. Can this be extended to multiple agents?
    
    \item Are there weaker or alternative forms of symmetry under which the principal can still achieve strictly better than a $1/2$-approximation in general?
\end{enumerate}

%%% Local Variables:
%%% mode: latex
%%% TeX-master: "main"
%%% End:

\section{Acknowledgments}
This paper is based upon work supported by the Air Force Office of Scientific Research under award number FA9550-24-1-0261, and the National Science Foundation under award number \mbox{CCF-2009060}. Any opinions, findings, and conclusions or recommendations expressed in this document are those of the authors and do not necessarily reflect the views of the United States Air Force or the National Science Foundation.

\bibliographystyle{ACM-Reference-Format}
\bibliography{stochastic}

\appendix
\section{Appendix}

\subsection{Single-Threshold Lower-Bound}
\label{proof:single-threshold-lower-bound}

Here we show that the threshold mechanisms of
\cref{thm:atomless-adversarial-lower-bound,,thm:adversarial-lower-bound,,thm:atomless-shuffled-adversarial-lower-bound,,thm:shuffled-adversarial-lower-bound}
% \cref{thm:atomless-adversarial-lower-bound}, \cref{thm:adversarial-lower-bound}, \cref{thm:atomless-shuffled-adversarial-lower-bound}, and \cref{thm:shuffled-adversarial-lower-bound}
all achieve an approximation of $1 - \Theta(\ln k / k)$.

\begin{proposition}
    Given an arbitrary integer $k > 0$, let $p$ be the unique positive real solution to $p^k + p - 1 = 0$. Then $p = 1 - \Theta\left(\frac{\ln k}{k}\right)$.
\end{proposition}

\begin{proof}
    Let $r = k (1 - p)$, so that $p = 1 - \frac{r}{k}$. Substituting into the polynomial, we get $g(r) = \left(1 - \frac{r}{k}\right)^k - \frac{r}{k}$. We aim to show that $g(r)$ has a root at $r = \Theta(\ln k)$. This would immediately yield the desired result of $p = 1 - \Theta\left(\frac{\ln k}{k}\right)$.
    
    Since $\left(1 - \frac{1}{t}\right)^t \le \frac{1}{e}$ for positive $t$, we have that $g(r) \le \exp(-r) - \frac{r}{k}$. If we choose to evaluate this at $r^- = 1 + \ln k$, it's easy to verify that $g(r^-) \le \frac{1}{e k} - \frac{1 + \ln k}{k} < 0$.
    
    Since $\left(1 - \frac{1}{t}\right)^{t - 1} \ge \frac{1}{e}$ for positive $t$, we have that $g(r) \ge \exp\left(\frac{-k}{k / r - 1}\right) - \frac{r}{k} = exp\left(\frac{-k r}{k - r}\right) - \frac{r}{k}$. If we choose to evaluate this at $r^+ = \frac{1}{2}(\ln k - \ln \ln k)$, then using the fact that $r^+ \le \frac{k}{2}$, we have $g(r^+) \ge \exp(-2 r) - \frac{r}{k} = \frac{\ln k}{k} - \frac{1}{2} \frac{\ln k - \ln \ln k}{k} > 0$.
    
    By the intermediate value theorem, polynomial $g(r)$ must have a root between $r^- = 1 + \ln k$ and $r^+ = \frac{1}{2}(\ln k - \ln \ln k)$. Therefore, $g(r)$ has a root at $\Theta(\ln k)$, completing the proof.
\end{proof}

\end{document}